\documentclass{article}
\usepackage{jheppub}

\usepackage{graphicx}

\usepackage{epsfig}

\usepackage{color}
\usepackage{amsmath}
\usepackage{type1cm}
\usepackage{xspace} 

\usepackage{ulem}

\usepackage{euscript}

\newcommand{\Li}[0]{\mathrm{Li}}

\title{
On the dependence of QCD splitting functions on the choice of the evolution
variable
} 

\author[a]{S.\ Jadach,}
\author[b]{A.~Kusina,}
\author[c]{W. P{\l}aczek,}
\author[a]{M.\ Skrzypek}
\affiliation[a]{Institute of Nuclear Physics, Polish Academy of Sciences,\\
                ul.\ Radzikowskiego 152, 31-342 Krak\'ow, Poland}
\affiliation[b]{Laboratoire de Physique Subatomique et de Cosmologie\\
                53 Rue des Martyrs Grenoble, France}
\affiliation[c]{Marian Smoluchowski Institute of Physics, Jagiellonian University,\\
                ul.\ {\L}ojasiewicza 11, 30-348 Krak\'ow, Poland}

\abstract{
We show that already at the NLO level the DGLAP evolution kernel $P_{qq}$ starts
to depend on the choice of the evolution variable. We give an explicit example of
such a variable, namely the maximum of transverse momenta of emitted partons
and we identify a class of evolution variables that leave the NLO
$P_{qq}$ kernel unchanged with respect to the known standard $\overline{\text{MS}}$ results.
The kernels are calculated
using a modified Curci--Furmanski--Petronzio method which is based on a direct
Feynman-graphs calculation.
}


\preprint{IFJPAN-IV-2016-3}

\begin{document}
\maketitle

\section{Introduction}

The choice of the evolution variable in the QCD evolution of the partonic
densities is one of the key issues in the construction of any Monte Carlo
parton shower \cite{Bengtsson:1986et}. The most popular choices are related to
virtuality, angle or transverse momentum of the emitted partons
\cite{Sjostrand:1985xi,Sjostrand:2004ef,Bahr:2008pv}. At the LO level, commonly used for the simulations, the
splitting functions are identical for all variables. In this note we
investigate whether it is the case also beyond the LO. To calculate the
evolution kernels we use slightly modified methodology of the
Curci-Furmanski-Petronzio classical paper
\cite{Curci:1980uw}. It is based on the direct calculation of the
contributing Feynman
graphs in the axial gauge, cf.\ \cite{Ellis:1978sf}. The graphs are extracted by
means of the projection
operators which close the fermionic or gluonic lines, put incoming partons
on-shell and extract pole parts of the expressions. The distinct
feature of this approach is the fact that the singularities are
regularized by means of the dimensional regularization, except for the ``spurious''
ones which are regulated by the PV (principal value)
prescription. To this end a dummy regulator $\delta$ is introduced with the help
of the replacement
\begin{align}
\frac{1}{ln} \to \frac{ln}{(ln)^2 +\delta^2(pn)^2}.
\end{align}
The regulator $\delta$ is
directly linked to the definition of the PV operation and has a simple
geometrical cut-off-like interpretation. This way some of the poles in
$\epsilon$ are replaced by logarithms of $\delta$. For more details we refer to
the original paper
\cite{Curci:1980uw} or to
later calculations, for example
\cite{Hei98,Jadach:2011kc,Gituliar:2014mua}. The
difference of our method with respect to the approach of \cite{Curci:1980uw}
is the use of the New PV (NPV) prescription which we have introduced in
\cite{Gituliar:2014eba,Skrzypek:2014tba}. It amounts to the
extension of the geometrical regularization to all singularities in the light
cone $l^+$ variable, not only to the ``spurious'' ones. This modification
turns out to be essential, as it further reduces the number of higher-order
poles in $\epsilon$ by replacing them with the $\log\delta$ terms,
and simplifies the contributions of individual graphs.

\begin{figure}
  \centerline{
    \includegraphics[height=2.65cm]{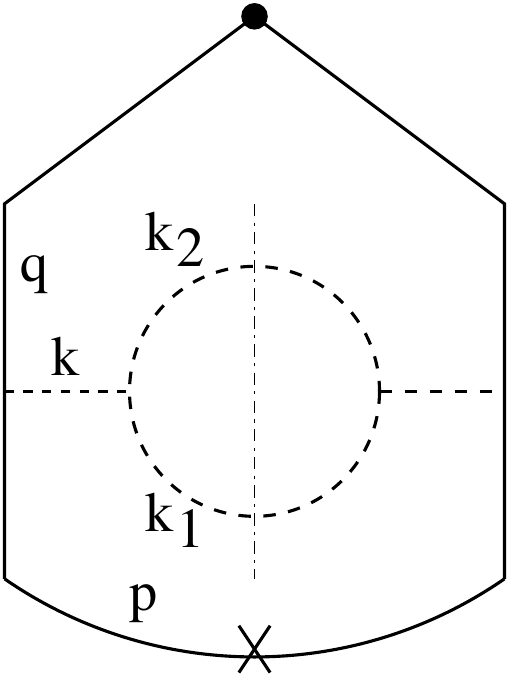}
    \hspace{1cm}
    \includegraphics[height=2.65cm]{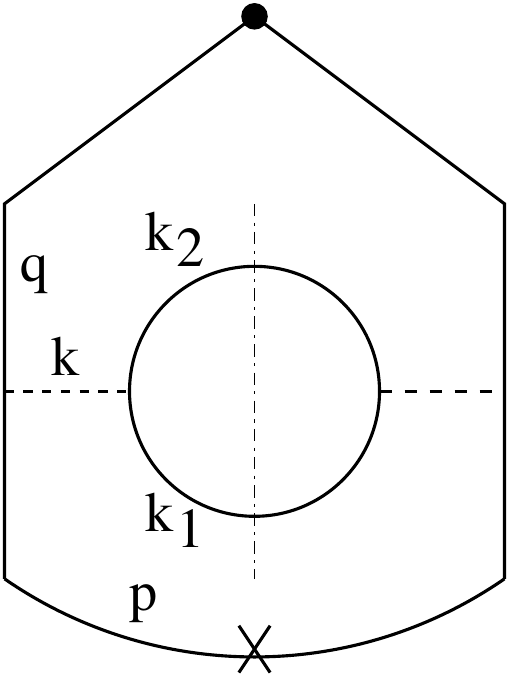}
    \hspace{1cm}
    \includegraphics[height=2.65cm]{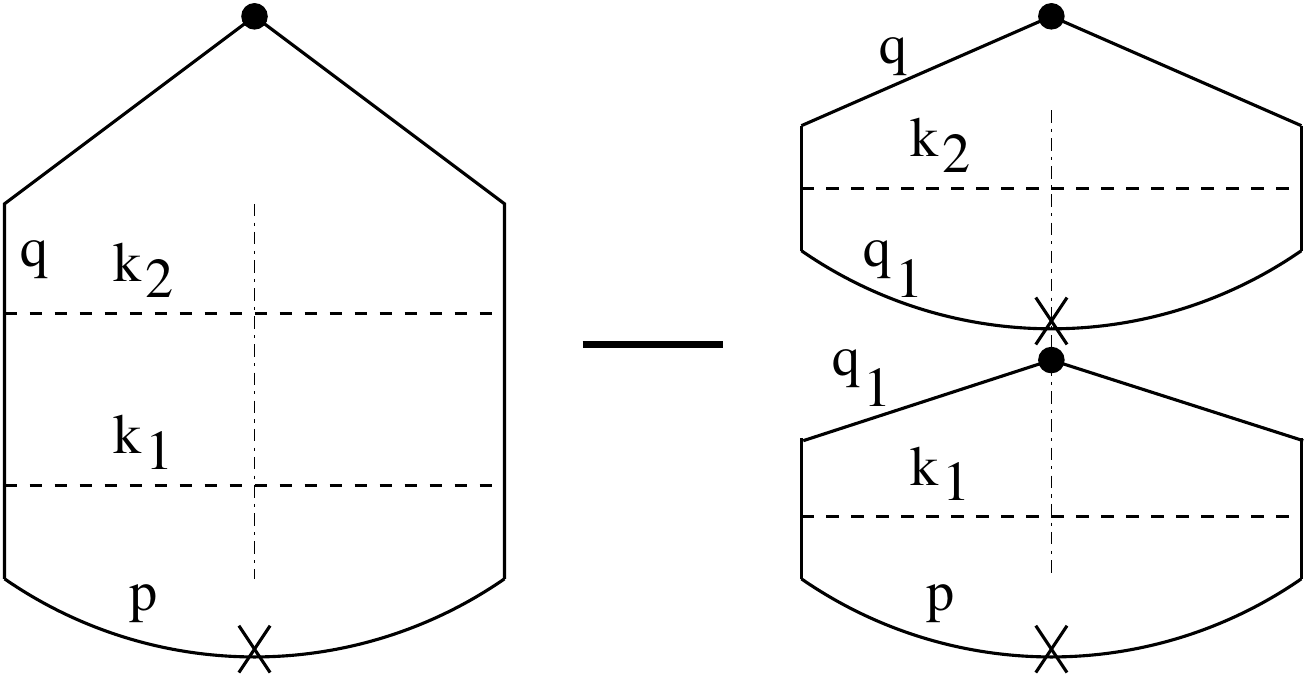}
  }
\centerline{     \hskip 0.7cm  Vg \hskip 2.7cm 
                  Vf \hskip 4cm 
                  Br$-$Ct  \hskip 2cm  }
\caption{\sf Real graphs with double poles contributing to the NLO non-singlet
$P_{qq}$
kernel.}
\label{fig:1}
\end{figure}

There are three mechanisms of keeping the kernel invariant under the change of
the cut-off: (1) Invariance of a particular diagram. This applies to all
diagrams with single poles in $\epsilon$. (2) Pairwise cancellation between
matching real and virtual graphs, as in the Vg and Vf graphs of Fig.~\ref{fig:1}.
(3) Cancellation between a graph and its counter-term. This is the case for ladder
graphs.
We will demonstrate that the mechanism (2) can fail already at the NLO level.

Our plan is the following. We will analyse the $P_{qq}$ kernel. There are three
graphs with second-order poles in $\epsilon$ contributing to the kernel. They
are depicted in Fig.\ \ref{fig:1}. 
We will calculate the difference between
the kernel with virtuality cut-off $-q^2<Q^2$, as in the original paper
\cite{Curci:1980uw},
and with the set of different cut-offs. The cut-offs we consider are:
the maximum and scalar sum of transverse momenta of the emitted
partons, i.e.\ $\max\{k_{1\perp},k_{2\perp}\}$ and $k_{1\perp}+k_{2\perp}$, as
well as
the maximum and total rapidity of the emitted partons, i.e.\
$\max\{k_{1\perp}/\alpha_1,k_{2\perp}/\alpha_2\}$ and
$|\vec k_{1\perp}+\vec k_{2\perp}|/(\alpha_1+\alpha_2)$.%
    \footnote{We define $k_{i\perp}\equiv |\vec k_{i\perp}|$.}
The calculation will
show that three of these cut-offs leave the kernel unchanged with respect to the
known standard $\overline{\text{MS}}$ results, whereas the
one on the maximum of transverse momenta, changes the kernel. We will show in
detail the mechanism of this change of the kernel and we will formulate a more
general scheme of its analysis.

\begin{figure}
  \centerline{
    \includegraphics[height=2.65cm]{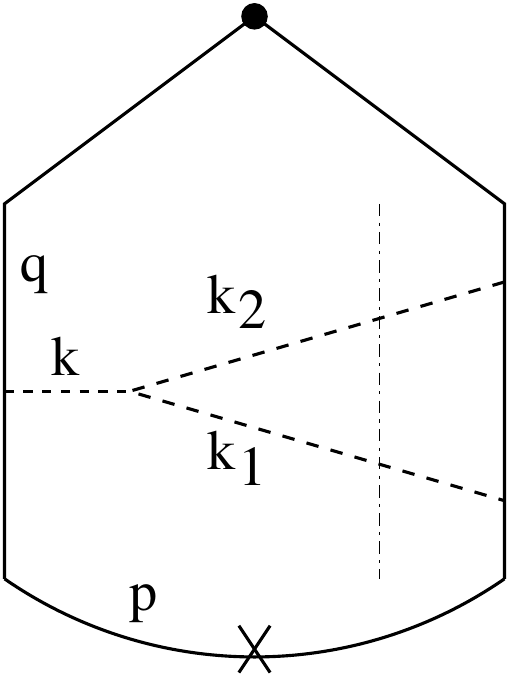}
  }
\caption{\sf The graph Yg contributing to the NLO non-singlet $P_{qq}$
kernel.}
\label{fig:2}
\end{figure}

We will start from the diagram named Vg and its sibling Vf. Next we will discuss
the ladder graph Br and its counter term, Ct. Our analysis will demonstrate that
only the Vg and Vf diagrams depend on the chosen cut-off variable. In the case of
the ladder graph the counter term cancels the dependence. Finally, we will
comment why the graphs with only single $\epsilon$ poles do not contribute. This
is also the reason why NPV is instrumental: it replaces $1/\epsilon^3$ poles of
the diagram Yg (depicted in Fig.\ \ref{fig:2}) 
by single poles and logarithms of
the regulator $\delta$. As a consequence, this diagram does not
contribute in NPV, whereas it would have a nontrivial contribution in the original
PV prescription.
\section{Diagram Vg}

In order to establish our notation and conventions, we give explicitly 
the starting formula for the contribution of the diagram Vg, corresponding to
Fig.\ \ref{fig:1}:
\begin{align}
\Gamma_{G} =& c_G^V g^4\;x\; \hbox{PP} \Biggl[\frac{1}{\mu^{4\epsilon}}\int
  d\Psi 
  \delta\Bigl(x-\frac{qn}{pn}\Bigr)
  \frac{1}{q^4}W_G
                     \Biggr],
\label{grG1}
\\
d\Psi =& 
  \frac{d^m k_1}{(2\pi)^m}2\pi \delta^+(k_1^2)
  \frac{d^m k_2}{(2\pi)^m}2\pi \delta^+(k_2^2)
=
        (2\pi)^{-2m+2} \frac{1}{4}
        \frac{d\alpha_1}{\alpha_1} \frac{d\alpha_2}{\alpha_2}
         d^{m-2}\vec k_{1\perp} d^{m-2}\vec k_{2\perp},
\label{grG2}
\\
c_G^V =& \frac{1}{2}C_G C_F,
\\
W_{G} =& \frac{1}{4qn} \frac{1}{k^4}\hbox{Tr} \Bigl(
  \hat n \hat q  \gamma^{\mu} \hat p \gamma^{\lambda} \hat q 
                                  \Bigr)
d_{\nu''\nu'}(k_2) d_{\mu\mu''}(k_1+k_2)
d_{\lambda''\mu'}(k_1) d_{\mu'\lambda}(k_1+k_2)
\notag
\\
   &\times V(k_1^{\mu''}+k_2^{\mu''},-k_2^{\nu''},-k_1^{\lambda''})
   V(k_1^{\mu'},k_2^{\nu'},-k_1^{\lambda'}-k_2^{\lambda'}).
\end{align}
We work in $m=4+2\epsilon$ dimensions. The Sudakov variables are defined with
the help of the light-like vector $n$ and the initial-quark momentum $p$:
\begin{align}
k_i & = \alpha_i p + \alpha_i^- n + k_{i\perp}^{(m)}, \;\;\;\;\;\;
q_i = x_i p + x^-_i n + q_{i\perp}^{(m)},
\end{align}
\begin{align}
p=(P,\vec{0},P),\;\;\; \;\;\;
n= \Bigl(\frac{pn}{2P},\vec{0},-\frac{pn}{2P}\Bigr).
\label{pin}
\end{align}
Note that the vector symbol $\vec{\phantom{m}}$ denotes ($m-2$)-dimensional
Euclidean vectors in transverse plane. 
Let us introduce new integration variables, $\vec \kappa_{1}$ and $\vec
\kappa_{2}$, instead of
 $\vec k_{1\perp}$ and  $\vec k_{2\perp}$:
\begin{align}
&\vec k_{1\perp} = \vec \kappa_{1} - \vec \kappa_{2},
\;\;\;\;\;\;\;\;
\vec k_{2\perp} = \frac{\alpha_2}{\alpha_1} \vec \kappa_{1} + \vec \kappa_{2},
\\
\hbox{i.e.}\;\;\; 
&\vec \kappa_{1} = \frac{\alpha_1}{\alpha_1+\alpha_2} \Bigl(\vec k_{1\perp}
+\vec k_{2\perp}\Bigr),
\;\;\;
\vec \kappa_{2} = \frac{\alpha_1\alpha_2}{\alpha_1+\alpha_2} 
                  \Bigl(\frac{\vec k_{2\perp}}{\alpha_2} - \frac{\vec
k_{1\perp}}{\alpha_1}\Bigr),
\label{kappy}
\end{align}
\begin{align}
\frac{\partial\vec k_{1\perp}\vec k_{2\perp}}
     {\partial\vec \kappa_1 \vec\kappa_2}
=&\;
\biggl(\frac{1-x}{\alpha_1}\biggr)^{m-2},
\end{align}
\begin{align}
d\Psi =&
        (2\pi)^{-2m+2} \frac{1}{4}
        \frac{d\alpha_1}{\alpha_1} \frac{d\alpha_2}{\alpha_2}
       \biggl(\frac{1-x}{\alpha_1}\biggr)^{m-2}
      \frac{1}{4} d\kappa_1^2 d\kappa_2^2
     d\Omega_{m-3}^{(1)}
     d\Omega_{m-3}^{(2)}
        \kappa_1^{m-4}
        \kappa_2^{m-4}.
\label{dPsik1k2}
\end{align}
The benefit of these variables is the diagonal form of the variables $k^2$ and
$q^2$ in which our formula is singular:
\begin{align}
k^2= \frac{(1-x)^2}{\alpha_1\alpha_2}\kappa_2^2,
\;\;\;\;
-q^2= \frac{1-x}{\alpha_1}\Bigl(
       \kappa_1^2\frac{1}{\alpha_1} +\kappa_2^2\frac{x}{\alpha_2}
                         \Bigr).
\label{kkqq}
\end{align}
The trace $W_G$ is of the form ($\theta$ is the angle between $\vec{\kappa_1}$
and
$\vec{\kappa_2}$)
\begin{align}
\label{WG2}
W_G=& \frac{8}{x(1-x)^2}
\biggl( \frac{\kappa_1^2}{\kappa_2^2} T_{Gc2} \cos^2\theta 
      + \sqrt{\frac{\kappa_1^2}{\kappa_2^2}} T_{Gc}\cos \theta 
      + \frac{\kappa_1^2}{\kappa_2^2} T_{GK} 
      + T_{Gn}
\biggr),
\\
T_{Gc2}= &4(1+\epsilon) \frac{x\alpha_2^2}{(1-x)^2},
\\
T_{Gc}= & x(1+x)\biggl(
         (1+\epsilon) 2(\alpha_1-\alpha_2)\frac{\alpha_2}{(1-x)^2}
        +\frac{\alpha_2-\alpha_1}{\alpha_1}
                 \biggr),
\\
T_{GK}= &\frac{\alpha_1^2+\alpha_2^2}{\alpha_1^2}
         \Bigl(1+x^2 +\epsilon(1-x)^2\Bigr)
     +\alpha_2^2(1+\epsilon),
\\
T_{Gn} = &(1+\epsilon)\frac{x^2}{(1-x)^2}(\alpha_1-\alpha_2)^2.
\end{align}
This allows us to rewrite formula (\ref{grG1}) as
\begin{align}
\Gamma_{G} =&\; c_G^V g^4\;x\; \hbox{PP}
\Biggl[\frac{1}{\mu^{4\epsilon}}\int
        (2\pi)^{-2m+2} \frac{1}{4}
        \frac{d\alpha_1}{\alpha_1} \frac{d\alpha_2}{\alpha_2}
       \biggl(\frac{1-x}{\alpha_1}\biggr)^{m-2}
\notag
\\ &\;\times
      \frac{1}{4} d\kappa_1^2 d\kappa_2^2
     d\Omega_{m-3}^{(1)}
     d\Omega_{m-3}^{(2)}
        \kappa_1^{m-4}
        \kappa_2^{m-4}
  \delta_{1-x-\alpha_1-\alpha_2}
\notag
\\
&\times
  \frac{1}{q^4}
  \frac{8}{x(1-x)^2}
\biggl( \frac{\kappa_1^2}{\kappa_2^2} T_{Gc2} \cos^2\theta 
      + \sqrt{\frac{\kappa_1^2}{\kappa_2^2}} T_{Gc}\cos \theta 
      + \frac{\kappa_1^2}{\kappa_2^2} T_{GK} 
      + T_{Gn}
\biggr)
                     \Biggr].
\label{grG1-kt}
\end{align}

\subsection{Cut-off on $\max\{k_{1\perp},k_{2\perp}\}< Q$}
\label{Vg-kt}
Let us now perform the calculation of the Vg graph with the cut-off on
the transverse momentum: $\max\{k_{1\perp},k_{2\perp}\}< Q$. 
This diagram has two
$\epsilon$-type singularities, related to $1/q^2$ and $1/\kappa_2^2\sim 1/k^2$.
The kernel
is constructed from the single-pole part of the diagram. Therefore, if we were
able to separate the part of the diagram containing a double pole, we could
considerably easier calculate the remaining single-pole part.
This can be done if we
calculate the difference between $\max\{k_{1\perp},k_{2\perp}\}< Q$ and the
standard virtuality-based cut-off $-q^2<Q^2$. This way we exclude the region of
double pole. In the leftover difference the $d\kappa_2^2$ integral has to
generate pole in $\epsilon$ and we can discard all terms finite in $\epsilon$.

We will compute
\begin{align}
\Delta\Gamma_{Vg}^{k_\perp- q}=\Gamma_{G}\left(\max\{k_{1\perp},k_{2\perp}\}< Q\right) -
\Gamma_{G}(-q^2<Q^2).
\end{align}
The $-q^2>  Q^2$ translates into (see eq.\ (\ref{kkqq}))
\begin{align}
-q^2=& c_1^2\kappa_1^2 + c_2^2\kappa_2^2 >  Q^2
\Rightarrow
\int\limits_0 d\kappa_2^2(\kappa_2^2)^{-1+\epsilon} \!\!\!\!\!\!
\int\limits_{(1/c_1)^2 Q^2 - (c_2/c_1)^2\kappa_2^2} \!\!\! d\kappa_1^2 
     \frac{(\kappa_1^2)^{1+\epsilon}}{(c_1^2\kappa_1^2 + c_2^2\kappa_2^2)^2},
\label{lim_kk}
\\
& c_1^2 = \frac{1-x}{\alpha_1^2},~~~c_2^2= \frac{(1-x)x}{\alpha_1\alpha_2}.
\end{align}
In eq.\ (\ref{lim_kk}) we have shown only the singular parts of the integrand.
The singularities of the integral are located at 
$k^2= \frac{(1-x)^2}{\alpha_1\alpha_2}\kappa_2^2=0$, i.e.\ at $\kappa_2=0$
and at
$-q^2=c_1^2\kappa_1^2 + c_2^2\kappa_2^2 = 0$ i.e.\ at $\kappa_1=\kappa_2=0$.
As we can see from (\ref{lim_kk}), the $q^2=0$ area is excluded due to
subtraction of the $\Gamma_{G}(-q^2<Q^2)$ which is available in the
literature \cite{Curci:1980uw, Jadach:2011kc}. The
external integrals over $d\alpha$ cannot contribute additional $1/\epsilon$
poles as they are regulated by the NPV prescription. This is one of
the two key ingredients of the calculation. Since we are interested in the pole
part of $\Delta\Gamma$, we can expand the $d\kappa_2$ integrand in a standard
way:
\begin{align}
\label{expkap2}
d\kappa_2^2(\kappa_2^2)^{-1+\epsilon}
=
d\kappa_2^2\frac{1}{\epsilon}\delta_{\kappa_2^2=0} 
  + {\cal{O}}(\epsilon^0).
\end{align}
This allows us to set $\kappa_2$ to zero in the rest
of the formula (\ref{grG1-kt}), both in the integrand and in the integration limits.
Furthermore, we can drop the terms $T_{Gc}$ and $T_{Gn}$ which do not have
singularities in $\kappa_2^2$. Finally, we can set $\epsilon$ to zero in the
remaining part of the formula.
Altogether we obtain
\begin{align}
\Delta\Gamma_{Vg}^{k_\perp- q}
=&\; c_G^V g^4\;x
        (2\pi)^{-6} \frac{1}{2}\frac{1}{\epsilon}
\frac{1}{x}
\;\hbox{PP}\Biggl[\int
        \frac{d\alpha_1}{\alpha_1^3} \frac{d\alpha_2}{\alpha_2}
       \frac{1}{c_1^4}
  \delta_{1-x-\alpha_1-\alpha_2}
\notag
\\
&\times
\int\limits_{(1/c_1)^2 Q^2}
       \frac{d\kappa_1^2}{\kappa_1^2}
\int
     d\Omega_{1}^{(1)}
     d\Omega_{1}^{(2)}
\bigl( T_{Gc2} \cos^2\theta 
      +  T_{GK}
\bigr)
                     \Biggr].
\label{grG1-kt-4d}
\end{align}
Next, we have to fix the upper limit of the $d\kappa_1$ integral. We have
\begin{align}
&\max\{k_{1\perp},k_{2\perp}\}< Q
\notag \\
&\Rightarrow \max\left\{|\vec\kappa_1 -\vec\kappa_2|,
          \left|\frac{\alpha_2}{\alpha_1}\vec\kappa_1 +\vec\kappa_2\right|\right\}< Q
\notag \\
&\Rightarrow |\vec\kappa_1 -\vec\kappa_2|<Q,~~~
             \left|\frac{\alpha_2}{\alpha_1}\vec\kappa_1 +\vec\kappa_2\right|< Q.
\label{kaplim}
\end{align}
We are interested in the limits for $\kappa_1$ at the point $\kappa_2=0$.
Immediately from eq.\ (\ref{kaplim}) we find
\begin{align}
\kappa_1 <Q,~~~\frac{\alpha_2}{\alpha_1}\kappa_1 < Q.
\end{align}
We have to discuss the
integration limits for both of the angular integrals as well. One of the angles
is trivial and covers the entire range $(0,2\pi)$, as the system has rotational
symmetry.
The other angle, $\theta$, between $\vec\kappa_1$
and $\vec\kappa_2$, has a non-trivial integration range, which depends on kappas
and alphas. However, there is a subspace where this angle is also unlimited. It
is given by the conditions
\begin{align}
\kappa_1 +\kappa_2 <Q,~~~ \frac{\alpha_2}{\alpha_1}\kappa_1 +\kappa_2 < Q.
\label{angnolim}
\end{align}
It just happens that in the limit $\kappa_2=0$ eq.\ (\ref{angnolim}) coincides
with the entire range of $\kappa_1$.
This way we find ($c_0=\alpha_2/\alpha_1$)
\begin{align}
\label{minQ}
\int\limits_{(1/c_1)^2Q^2}^{\min\{Q^2/c_0^2, Q^2\}}&
\frac{d\kappa_1^2}{\kappa_1^2}
\int\limits_0^{2\pi}d\Omega_1^{(1)}
\int\limits_0^{2\pi}d\theta
\bigl( T_{Gc2} \cos^2\theta 
      +  T_{GK}
\bigr)
\\
=&\;
\biggl(\theta_{c_0<1}\ln{c_1^2} 
+
\theta_{c_0>1}\ln\frac{c_1^2}{c_0^2} \biggr) 2\pi 
\bigl( \pi T_{Gc2} 
      +  2\pi T_{GK}
\bigr)
\notag \\
=&\;
\Bigl(\theta_{\alpha_2<\alpha_1}\ln\frac{1-x}{\alpha_1^2}
+
\theta_{\alpha_2>\alpha_1}\ln\frac{1-x}{\alpha_2^2}\Bigr) 4\pi^2
\frac{\alpha_2}{\alpha_1 x} T_S,
\notag \\
T_S=& x(1+x^2)\biggl(\frac{1}{(1-x)^2}\alpha_1\alpha_2 
           +\frac{\alpha_1^2+\alpha_2^2}{\alpha_1\alpha_2}
      \biggr).
\end{align}
Going back to eq.\ (\ref{grG1-kt-4d}) we obtain
\begin{align}
\Delta\Gamma_{Vg}^{k_\perp- q} =&\; c_G^V \frac{g^4}{(2\pi)^{4}} 
\frac{1}{2\epsilon}
\frac{1}{x(1-x)^2}
\;\int {d\alpha_1}{d\alpha_2}
\delta_{1-x-\alpha_1-\alpha_2}
\biggl(\ln(1-x)
-  4 \theta_{\alpha_2<\alpha_1}\ln\alpha_1 \biggr)
 T_S.
\label{grG1-kt-4d-2}
\end{align}
Performing the $\alpha$-integrals we find
\begin{align}
\Delta\Gamma_{Vg}^{k_\perp- q} =& \; 
c_G^V \Bigl(\frac{\alpha_S}{\pi}\Bigr)^2 
\frac{1}{2\epsilon}
\frac{1+x^2}{1-x}
\biggl[\ln\frac{1}{(1-x)} 
       \Bigl(2I_0+2\ln (1-x)-\frac{11}{6}\Bigr)
-  4 \biggl(
-\frac{11}{12}\ln 2
+\frac{131}{144}
-\frac{\pi^2}{12}\biggr)
\biggr]
\label{DGstrange}
\end{align}
where we have introduced the symbol $I_0$ for the IR-divergent integral regularized by
means of the PV prescription with the geometrical $\delta$ parameter:
\begin{align}
I_0= &
\int_0^{1}
        d\alpha
\frac{\alpha}{\alpha^2+\delta^2}
=
-\frac{1}{2}\ln\delta^2,
\\
I_1=& 
\int_0^{1}
        d\alpha \ln \alpha
\frac{\alpha}{\alpha^2+\delta^2}
=
-\frac{1}{8}\ln^2\delta^2 -\frac{\pi^2}{24}.
\label{I0I1}
\end{align}
The result (\ref{DGstrange}) differs from the shift in virtual corrections
that we show later in Section~\ref{virtual}.
We have obtained a net change of the kernel.
\subsection{Cut-off on $k_{1\perp}+k_{2\perp}< Q$}
\label{sect:pos}
\begin{figure}
  \centerline{
    \includegraphics[height=2.65cm]{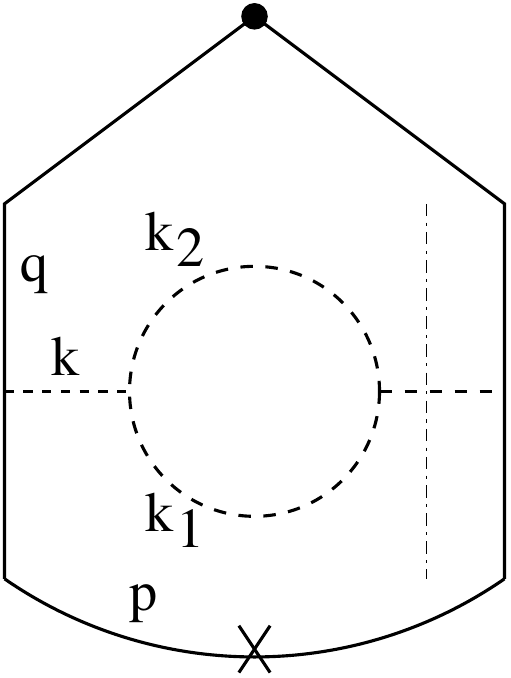}
  }
\caption{\sf Real-virtual graph Vg contributing to NLO non-singlet $P_{qq}$
kernel}
\label{fig:3}
\end{figure}

We have demonstrated in the previous section that the change of real and virtual
Vg-type diagrams do not compensate each other. Why is this so? Imagine the
virtual correction Vg, Fig.\ \ref{fig:3}.
The graph has one real gluon, labelled
$k$, and the cut-off is unique and trivial:
$
k_\perp \leq Q
$.
However, if we look inside the graph we find two virtual momenta, $k_1$ and
$k_2$, such that $k_1 +k_2=k$. Therefore, our $k_\perp$-cut-off at the
unintegrated level is $|\vec k_{1\perp}+ \vec k_{2\perp}|\leq Q$.
This cut-off
seems to be not good for real gluons because it does not close the phase space.
We will come to this issue in the next paragraph. For now let us note that,
as argued in Section \ref{Vg-kt}, we calculate only the difference
between the $q^2$ and $k_\perp$ cut-offs, and therefore we integrate only 
over the region singular in $\kappa_2$, i.e.\ we expand the $d\kappa_2$
integral according to eq.\ (\ref{expkap2}). This introduces $\kappa_2^2=
[\alpha_1\alpha_2/(1-x)^2]k^2=0$, or equivalently $\vec k_{1\perp}/\alpha_1-
\vec k_{2\perp}/\alpha_2=0$. In this subspace the condition $|\vec k_{1\perp}+
\vec k_{2\perp}|\leq Q$ simplifies to 
$\kappa_1^2\leq [\alpha_1/(1-x)]^2 Q^2 =[1/(1+c_0)]^2Q^2$. In analogy, the
``scalar" condition $|\vec k_{1\perp}|+
|\vec k_{2\perp}|\leq Q$ simplifies to 
$|\vec\kappa_1|+|\vec\kappa_1|(\alpha_2/\alpha_1)\leq Q$,
i.e.\
$\kappa_1^2\leq [\alpha_1/(1-x)]^2 Q^2$, identical to the previous cut-off.
Therefore, we expect that the ``scalar'' cut-off $|\vec k_{1\perp}|+
|\vec k_{2\perp}|\leq Q$ will give the result compatible with
the virtual correction.
With this cut-off eq.\ (\ref{minQ})
becomes
\begin{align}
\label{minQbetter}
\int\limits_{(1/c_1)^2Q^2}^{[1/(1+c_0)]^2Q^2}
& \frac{d\kappa_1^2}{\kappa_1^2}
\int\limits_0^{2\pi}d\Omega_1^{(1)}
\int\limits_0^{2\pi}d\theta
\bigl( T_{Gc2} \cos^2\theta 
      +  T_{GK}
\bigr)
\\
=&
\ln\frac{c_1^2}{(1+c_0)^2} 2\pi 
\bigl( \pi T_{Gc2} 
      +  2\pi T_{GK}
\bigr)
\notag\\
=&
\notag
\ln\frac{1}{1-x} 4\pi^2 \frac{\alpha_2}{\alpha_1 x} T_S.
\end{align}
Consequently, eq.\ (\ref{grG1-kt-4d-2}) becomes
\begin{align}
\Delta\Gamma_{Vg}^{\Sigma k_\perp -q} =& c_G^V
\frac{g^4}{(2\pi)^{4}} 
\frac{1}{2\epsilon}
\frac{1}{x(1-x)^2}
\;\int {d\alpha_1}{d\alpha_2}
\delta_{1-x-\alpha_1-\alpha_2}
\ln\frac{1}{1-x} T_S
\label{grG1-kt-4d-2-better}
\\
=& c_G^V \Bigl(\frac{\alpha_S}{\pi}\Bigr)^2
\frac{1}{2\epsilon}
\frac{1+x^2}{1-x}
       \ln\frac{1}{1-x} \Bigl(2I_0+2\ln (1-x)-\frac{11}{6}\Bigr).
\label{DGOK}
\end{align}
This way we reproduced result (\ref{DGstrange}), but without additional constant
terms. It is identical to the change in virtual corrections and there is no
modification of the kernel.
\subsection{Cut-off on $|\vec k_{1\perp}+ \vec k_{2\perp}|\leq Q$}
\label{sect:ktcut}
Let us come back to the cut-off on vector variable $|\vec k_{1\perp}+ \vec
k_{2\perp}|\leq Q$. It indeed allows for arbitrarily big $\vec
k_{i\perp}$ vectors. The question is however whether it leads to
well-defined and meaningful kernels? We will argue that it does.

Translated
into $\kappa$-variables of eq.\ (\ref{kappy}), the cut-off is simply 
$\kappa_1\leq \alpha_1/(1-x) Q$, identical to the one of Section
\ref{sect:pos}. The $\vec\kappa_2 = \vec\kappa_1-\vec k_{1\perp}$ variable is
unbounded because so is $\vec k_{1\perp}$ (the $\vec k_{2\perp}$ can always be
adjusted to fulfill the cut-off) and the angle is also unlimited, $0\leq
\angle(\vec\kappa_1,\vec\kappa_2)\leq 2\pi$. Keeping in mind the
discussion on the origin of poles given around eq.\ (\ref{expkap2}),
we conclude that the upper limit on $\kappa_2$ does not matter at
all, and we can set it to infinity as well. Repeating all the steps of Section
\ref{sect:pos} we recover the result (\ref{DGOK}). In other
words, we have just shown that the cut-off $|\vec k_{1\perp}+ \vec
k_{2\perp}|\leq Q$ leads to a proper kernel.

One may be worried about the higher order terms of $\epsilon$-expansion of eq.\
(\ref{expkap2}): are they finite? To answer this question let us inspect the
original equations (\ref{grG1}) and (\ref{WG2}). In the limit
$\kappa_2^2\to\infty$ we have $-q^2\sim (1-x)x/(\alpha_1\alpha_2)\kappa_2^2$
and we find integrals of the type
\begin{align}
\int\limits^\infty d\kappa_2^2 
\left\{
\frac{1}{(\kappa_2^2)^3},\frac{1}{(\kappa_2^2)^{5/2}},\frac{1}{(\kappa_2^2)^2}
\right\},
\end{align}
which are integrable at the infinity. We conclude that the $\epsilon$ expansion
of eq.\ (\ref{expkap2}) is legitimate and the cut-off $|\vec k_{1\perp}+ \vec
k_{2\perp}|\leq Q$ is self consistent. The open question is though how will
this cut-off perform with other graphs. Another question concerns its
generalization to more than two real partons.

\subsection{Cut-off on rapidity}
Let us briefly  comment on the cut-off on rapidity. By rapidity we understand
the quantity $a=|\vec k_{\perp}|/\alpha$ (massless) or $a=\sqrt{|\vec
k_{\perp}|^2+k^2}/\alpha$ (massive). For the case of two emissions the
analogy to virtual graph leads to
 $a=|\vec k_{1\perp} +\vec k_{2\perp}|/(\alpha_1+\alpha_2)\leq Q$ or
 $a=\sqrt{|\vec k_{1\perp} +\vec k_{2\perp}|^2+(k_1+k_2)^2}
/(\alpha_1+\alpha_2)\leq Q$. In the subspace $\kappa_2^2\sim k^2=0$ both
formulas coincide and both are identical to the $k_{\perp}$-type formula with
the cut-off $Q$ shifted to $Q(1-x)$ in the
$k_{\perp}$-type formula. This is just the result we have obtained for the virtual
corrections. Another option is $\max\{a_1,a_2\}\leq Q$. We have 
$\vec a_1=(\vec\kappa_1-\vec\kappa_2)/\alpha_1$ and 
$\vec a_2=\vec\kappa_1/\alpha_1+\vec\kappa_2/\alpha_2$. At $\kappa_2=0$ this
leads to $\kappa_1/\alpha_1\leq Q$ or equivalently 
$|\vec k_{1\perp} +\vec k_{2\perp}|/(\alpha_1+\alpha_2)\leq Q$. This is
identical to the previous case, so we expect the result to be in agreement with the virtual
correction as well.

Let us compute the correction from the $q^2$-type to $a$-type cut-off.
To this end we generalize eq.\ (\ref{minQ}), which is the $k_\perp$-type, by
replacing $Q^2\to Q^2(1-x)^\sigma$ in the upper limit: $\sigma=2$
corresponds to the rapidity case discussed here, $\sigma=0$ is the $k_\perp$
case (reference) and $\sigma=1$ is the virtuality case (the correction vanishes).
This is so because:
$\Sigma k_\perp=((1-x)/\alpha_1)^2\kappa_1^2\leq Q^2$ is described by eq.\
(\ref{minQ}). $a=k_\perp/(1-x) \to \kappa_1/\alpha_1 \leq Q$ requires
multiplication of $Q^2$ by $(1-x)^2$ (with respect to the $k_\perp$ case).
$-q^2\to  (1-x)\kappa_1^2/\alpha_1^2 \leq Q^2$ requires
multiplication of $Q^2$ by $1-x$.
\begin{align}
\label{minQbettersigma}
\int\limits_{(1/c_1)^2Q^2}^{[(1-x)^\sigma/(1+c_0)]^2Q^2}
& \frac{d\kappa_1^2}{\kappa_1^2}
\int\limits_0^{2\pi}d\Omega_1^{(1)}
\int\limits_0^{2\pi}d\theta
\bigl( T_{Gc2} \cos^2\theta 
      +  T_{GK}
\bigr)
\\
=&
\ln\frac{c_1^2(1-x)^\sigma}{(1+c_0)^2} 2\pi 
\bigl( \pi T_{Gc2} 
      +  2\pi T_{GK}
\bigr)
\notag
\\
\notag
=&
\ln{(1-x)^{\sigma-1}} 4\pi^2 \frac{\alpha_2}{\alpha_1 x} T_S.
\end{align}
Consequently, eq.\ (\ref{grG1-kt-4d-2}) becomes
\begin{align}
\Delta\Gamma_{Vg}^{\sigma -q} =& c_G^V \frac{g^4}{(2\pi)^{4}} 
\frac{1}{2\epsilon}
\frac{1}{x(1-x)^2}
\;\int {d\alpha_1}{d\alpha_2}
\delta_{1-x-\alpha_1-\alpha_2}
\ln{(1-x)^{\sigma-1}} T_S
\label{grG1-kt-4d-2-bettersigma}
\notag\\
=& c_G^V \Bigl(\frac{\alpha_S}{\pi}\Bigr)^2
\frac{1}{2\epsilon}
\frac{1+x^2}{1-x}
       \ln{(1-x)^{\sigma-1}} \Bigl(2I_0+2\ln
(1-x)-\frac{11}{6}\Bigr).
\end{align}

\subsection{General rule}
\label{gen-rule}

We can now generalize the analysis of previous sections and formulate a more
universal rule of identifying the variables that do or do not change the
NLO kernel.

\begin{figure}[!ht]
  \centering
  {\epsfig{file=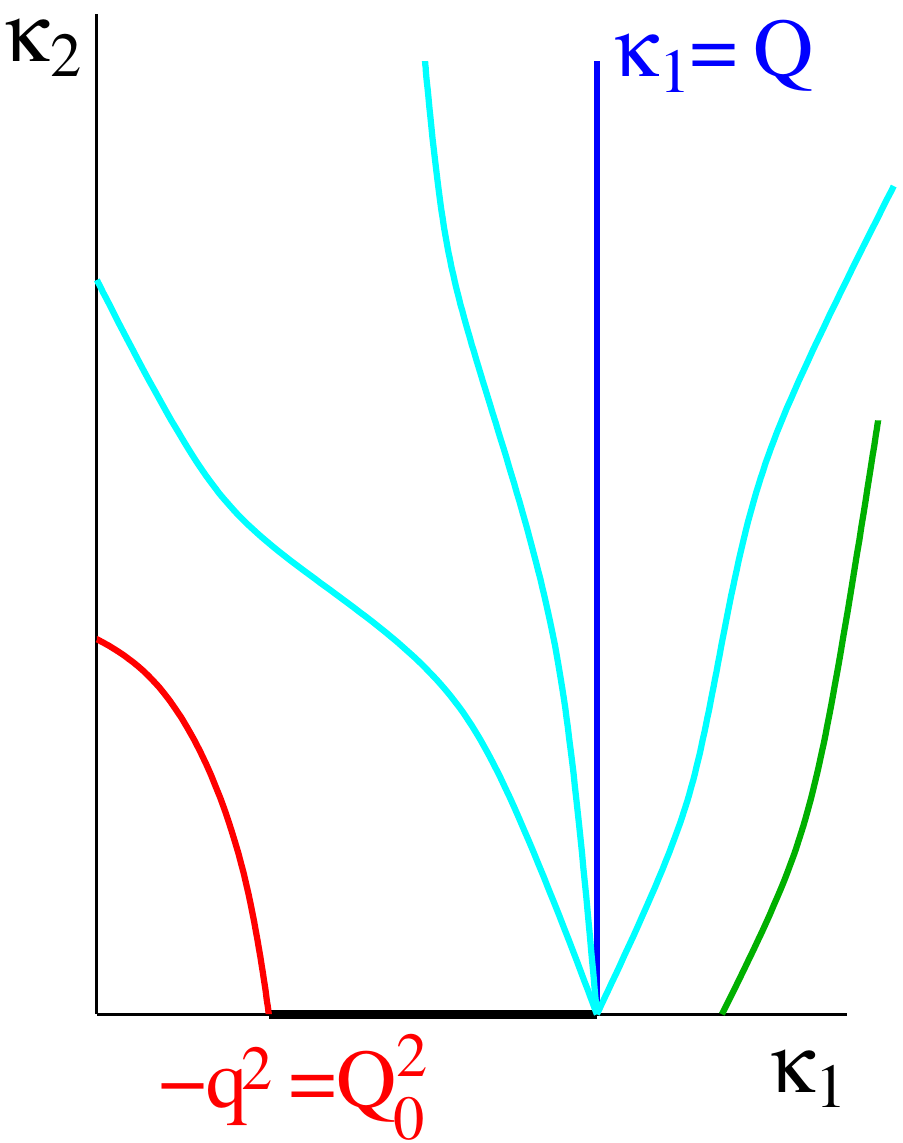,width=40mm}}
  \caption{\sf 
The $(\kappa_1,\kappa_2)$ plane. The
cut-off $\kappa_1\leq Q$ is shown in dark blue. A family of
other cut-off lines is shown in light blue. At the bottom left
the $-q^2\leq Q_0^2$ line is plotted in red. The singularities lie at the origin
of the frame ($q^2=0$) and along the line $\kappa_2^2\sim k^2=0$.
The integration path is the thick black line along $\kappa_2=0$ between the crossing
points of $-q^2=Q_0^2$ and the cut-off with the axis.
}
  \label{fig:kap1kap2}
\end{figure}

In Fig.~\ref{fig:kap1kap2} we show the $(\kappa_1,\kappa_2)$ plane. The
blue cut-off $\vec\kappa_1\leq Q$ is shown along with a family of other
cut-off lines. Some of them (blue) are equivalent if they cross
the $\kappa_1$-axis at the same point. The cut-offs may close the
$\kappa_2$-direction from above or leave it open. At the bottom left we plot
the red $-q^2\leq Q_0^2$ line. The singularities lie at the origin of the
frame ($q^2=0$) and along the line $\kappa_2^2\sim k^2=0$.
The integration path is the thick line along $\kappa_2=0$ between crossing
points of $-q^2=Q_0^2$ and the cut-off with the axis.

The strategy we use is the following. We take a group of 
variables that coincide at the LO level (i.e.\ for single emission), we
express them
in terms of the variables $\kappa$ and we set $\kappa_2=0$. All the variables that
cross the $\kappa_1$ axis at the same point will lead to the same result. It is
now a matter of choosing one of them, calculating the shift, as outlined in the
paper, and comparing it with the shift in the virtual corrections. We collect the
shifts in the virtual corrections for the basic three types of variables in Section~\ref{virtual}.

\section{Diagram Vf}
\label{Vf-parameterizations}
Let us now perform the analysis of the Vf graph. It will heavily rely on the
analysis done for the Vg graph. Let us begin with the $\max\{k_{1\perp},k_{2\perp}\}$
calculation.
Our starting point is the diagram depicted in Fig.~\ref{fig:1}. The analytical formula
is analogous to eq.\ (\ref{grG1}): 
\begin{align}
\Gamma_{F} =&\; c_F^V g^4\;x\; \hbox{PP} \Biggl[\frac{1}{\mu^{4\epsilon}}\int
  d\Psi 
  \delta\Bigl(x-\frac{qn}{pn}\Bigr)
  \frac{1}{q^4}W_F
                     \Biggr],
\label{grF1}
\\
c_F^V =&\; C_FT_F,
\\
W_{F} =&\; \frac{1}{4qn} \frac{1}{k^4}\hbox{Tr} \Bigl(
  \hat n \hat q  \gamma^{\mu} \hat p \gamma^{\lambda} \hat q 
                                  \Bigr)
d_{\mu\mu''}(k_1+k_2)
\hbox{Tr} \Bigl( \hat k_2  \gamma^{\mu''} \hat k_1 \gamma^{\mu'}
          \Bigr)
d_{\mu'\lambda}(k_1+k_2)
\notag\\
\label{WF2}
=&\; \frac{32pn}{4qn}\frac{1}{(1-x)^2}
\biggl( \frac{\kappa_1^2}{\kappa_2^2} T_{Fc2} \cos^2\theta 
      + \sqrt{\frac{\kappa_1^2}{\kappa_2^2}} T_{Fc}\cos \theta 
      + \frac{\kappa_1^2}{\kappa_2^2} T_{FK} 
      + T_{Fn}
\biggr),
\\
T_{Fc2}= &\; -4x \frac{\alpha_2^2}{v^2},
\\
T_{Fc}= &\; 2x(1+x)\alpha_2
         (\alpha_2-\alpha_1)\frac{1}{v^2},
\\
T_{FK}= &\; \frac{1}{2}\epsilon v^2\frac{\alpha_2}{\alpha_1}
        +\frac{1}{2}(1+x^2)\frac{\alpha_2}{\alpha_1}
        -\alpha_2^2,
\\
T_{Fn} = &\; 4\frac{x^2}{v^2}\alpha_1\alpha_2.
\end{align}
The calculation goes now in a complete analogy to the Vg case and we arrive 
at the adapted version of eq.\ (\ref{grG1-kt-4d-2}) into which we plug
in the $T_S^{(F)}$ function 
\begin{align}
\Delta\Gamma_{Vf}^{k_\perp -q} =&\; c_F^V \frac{g^4}{(2\pi)^{4}} 
\frac{1}{2\epsilon}
\frac{1}{x(1-x)^2}
\;\int {d\alpha_1}{d\alpha_2}
\delta_{1-x-\alpha_1-\alpha_2}
\biggl(\ln(1-x)
-  4 \theta_{\alpha_2<\alpha_1}\ln\alpha_1 \biggr)
 T_S^{(F)},
\label{grF1-kt-4d-2}
\\
T_S^{(F)}=&\;\frac{\alpha_1}{\alpha_2}x
         \biggl(\frac{1}{2}T_{Fc2}^{(0)} + T_{FK}^{(0)}\biggr)
   = \frac{1}{2}x(1+x^2)\biggl(-2\frac{1}{(1-x)^2}\alpha_1\alpha_2 
           +1
      \biggr).
\end{align}
Once the $d\alpha$-integration is done we obtain the final result for the
Vf graph with the cut-off on $\max k_\perp$
\begin{align}
\Delta\Gamma_{Vf}^{k_\perp -q}=
c_F^V \Bigl(\frac{\alpha}{2\pi}\Bigr)^2 
\frac{2}{\epsilon}
\frac{1+x^2}{1-x}
\biggl[
-\frac{1}{3}\ln(1-x)
  +\frac{23}{36}  -\frac{2}{3}\ln 2 
\biggr].
\label{VfkTq}
\end{align}
Let us discuss also the other choices of the cut-offs: the sum of $k_\perp$, virtuality
and rapidity, labelled as $\sigma =0,1,2$, respectively. For this purpose it is enough
to repeat the analysis and reuse the formulas for the Vg graph.
The formula (\ref{grG1-kt-4d-2-bettersigma}) can be directly used to give
\begin{align}
\Delta\Gamma_{Vf}^{\sigma -q} =&\; c_F^V \frac{g^4}{(2\pi)^{4}} 
\frac{1}{2\epsilon}
\frac{1}{x(1-x)^2}
\;\int {d\alpha_1}{d\alpha_2}
\delta_{1-x-\alpha_1-\alpha_2}
\ln{(1-x)^{\sigma-1}} T_S^{(F)}
\notag \\
\label{grG1-kt-4d-2-bettersigmaF}
=&\; c_F^V \Bigl(\frac{\alpha_S}{2\pi}\Bigr)^2
\frac{2}{\epsilon}
\frac{1+x^2}{1-x}
       \frac{1}{3}\ln{(1-x)^{\sigma-1}}.
\end{align}
\begin{align}
\Delta\Gamma_{Vf}^{\Sigma k_\perp -q}=&\;
c_F^V \Bigl(\frac{\alpha}{2\pi}\Bigr)^2 
\frac{2}{\epsilon}
\frac{1+x^2}{1-x}
\biggl[
-\frac{1}{3}\ln(1-x)
\biggr],
\\
\Delta\Gamma_{Vf}^{a -q}=&\;
c_F^V \Bigl(\frac{\alpha}{2\pi}\Bigr)^2 
\frac{2}{\epsilon}
\frac{1+x^2}{1-x}
\biggl[
\frac{1}{3}\ln(1-x)
\biggr].
\end{align}
\section{Virtual diagrams}
\label{virtual}
The shift in virtual corrections due to change of the cut-off can be found in
Ref.\ \cite{Gituliar:2014mua}. The $\sigma$-dependence of each
diagram is given
there. One finds that there is no $\sigma$-dependence for the $C_F^2$-type
graphs and the only ones that do depend on $\sigma$ are Vg and Vf, see eqs.\ (4.25) and (4.31)
in Ref.\ \cite{Gituliar:2014mua}. Here we quote the
change with respect
to the virtuality case:
\\
\begin{align}
\Delta\Gamma_{virt}^{\sigma -q}=
\Bigl( \frac{\alpha}{2\pi}\Bigr)^2\frac{1}{2\epsilon}
C_F \frac{1+x^2}{1-x}
 \Bigl( \beta_0 -4 C_A \bigl(I_0 +\ln (1-x)\bigr) \Bigr) \ln^{\sigma-1}(1-x).
\end{align}
\section{Combined Vg$+$Vf real diagrams}
Let us combine the Vg and Vf real graphs for the case of $\max\{k_{1\perp},k_{2\perp}\}$.
The formulas to be added are (\ref{DGstrange}) and (\ref{VfkTq}) with $c_G^V =
(1/2) C_FC_A$ and $c_F^V = C_F T_F$:
\begin{align}
\Delta\Gamma_{Vf+Vg}^{k_\perp -q} =&\; 
 C_F \Bigl(\frac{\alpha_S}{2\pi}\Bigr)^2 
\frac{2}{\epsilon} \frac{1+x^2}{1-x}
\biggl[ -C_A\Bigl(I_0+\ln(1-x)\Bigr)\ln(1-x)
\notag
\\&
+C_A\frac{\pi^2}{6} -C_A\frac{1}{16}
     +\frac{1}{4}\beta_0 \ln(1-x)
+ \frac{1}{2}\beta_0\ln 2
-\frac{23}{48}\beta_0  
\biggr],
\label{VfkTqq}
\\ 
\beta_0= &\; \frac{11}{3}C_A-\frac{4}{3}T_F.
\end{align}
Anticipating
the results of the following sections we can state that this result represents the
change of the $P_{qq}$ kernel due to the real corrections when the evolution
variable (cut-off) is changed from the standard $q^2$ one to $\max\{k_{1\perp},k_{2\perp}\}$.
Supplied with the virtual corrections it will give the complete effect.

Let us combine also the $\sigma$-type cut-offs for the real Vf$+$Vg graphs
\begin{align}
\Delta\Gamma_{Vf+Vg}^{\sigma -q} =& \; C_F\Bigl(\frac{\alpha_S}{2\pi}\Bigr)^2
\frac{1}{2\epsilon}
\frac{1+x^2}{1-x} \ln{(1-x)^{\sigma-1}} 
        \Bigl[-\beta_0 +4C_A\Bigl(I_0+\ln (1-x)\Bigr)\Bigr].
\end{align}

\section{Added real and virtual diagrams}
We can now add changes of the real and virtual Vf$+$Vg graphs. For the $\sigma$-type
cut-offs we observe that the contributions cancel each other and there is no net
effect, as expected. The situation is different for the cut-off on
$\max\{k_{1\perp},k_{2\perp}\}$, where we find the following shift
\begin{align}
\Delta\Gamma_{Vf+Vg,R+V}^{k_\perp -q} =
 C_F \Bigl(\frac{\alpha_S}{2\pi}\Bigr)^2 
\frac{1}{2\epsilon} \frac{1+x^2}{1-x}
\biggl[ 
C_A\frac{2\pi^2}{3} -C_A\frac{1}{4}
+ 2\beta_0\ln 2
-\frac{23}{12}\beta_0  
\biggr].
\end{align}
 This result can be translated into the kernel $P_{qq}$ which is the residue of
$\Gamma$ \cite{Curci:1980uw}:
\begin{align}
\Gamma =&\; \delta_{1-x} +\frac{1}{\epsilon}\Bigl[
    \Bigl(\frac{\alpha}{2\pi}\Bigr) P^{(1)}
   +\frac{1}{2}\Bigl(\frac{\alpha}{2\pi}\Bigr)^2 P^{(2)} + \dots \Bigr],
\\
P_{qq} = &\; \Bigl(\frac{\alpha}{2\pi}\Bigr) P^{(1)}
   +\Bigl(\frac{\alpha}{2\pi}\Bigr)^2 P^{(2)}+ \dots,
\end{align}
and we obtain the following change of the $P_{qq}$ kernel
\begin{align}
P_{qq}&(\max\{k_{1\perp},k_{2\perp}\}< Q) -P_{qq}(-q^2<Q^2) =
\notag
\\
= &\;
 C_F \Bigl(\frac{\alpha_S}{2\pi}\Bigr)^2 
\frac{1+x^2}{1-x}
\biggl[ 
C_A\Bigl( \frac{2\pi^2}{3} -\frac{1}{4}\Bigr)
+ \beta_0\Bigl(2\ln 2 -\frac{23}{12}\Bigr)  
\biggr].
\end{align}
This is the central new result of this paper.
\section{Br (ladder) graph and counter term}
\label{sect:ladder-other}
We turn now to the ladder graph and a counter term associated with it, shown in
Fig.~\ref{fig:1}. Both of
them have double $\epsilon$ poles and therefore can be modified once the evolution
variable changes. However, we will demonstrate now that their difference
remains unchanged. 

The contribution $\Gamma_{Br}$ of the ladder graph is similar to the one given
for the Vg graph in eqs.\
(\ref{grG1}, \ref{grG2})
\begin{align}
\Gamma_{Br} =&\; C_F^2 \frac{g^4\;x}{(2\pi)^6} \hbox{PP} \Biggl[
  \frac{(2\pi)^{-2\epsilon}}{\mu^{2\epsilon}}
  \int
        \frac{d\alpha_2}{2\alpha_2}
         d^{2+2\epsilon}\vec k_{2\perp} 
\frac{(2\pi)^{-2\epsilon}}{\mu^{2\epsilon}}
  \int \frac{d\alpha_1}{2\alpha_1}  d^{2+2\epsilon}\vec k_{1\perp}
  \delta_{1-x-\alpha_1-\alpha_2}
  \frac{1}{q^4}
  \frac{1}{q_1^4}
       W_{Br} \Biggr].
\label{ladderA}
\end{align}
\begin{align}
W_{Br} &= \frac{1}{4qn} \hbox{Tr} \Bigl(
  \hat n \hat q \hat \gamma^\mu \hat q_1 \hat \gamma^\alpha \hat p
  \hat \gamma^\beta \hat q_1 \hat \gamma^\nu \hat q
           \Bigr) d_{\alpha\beta}(k_1) d_{\mu\nu}(k_2).
\\
&= \frac{4}{x\alpha_1\alpha_2}
                   \frac{k_{1\perp}^2}{\alpha_1}
    \Bigl(
    \frac{k_{1\perp}^2}{\alpha_1} T_1  
  + \frac{k_{2\perp}^2}{\alpha_2} T_2 
  + 2\vec k_{1\perp}\cdot \vec k_{2\perp} T_3
    \Bigr),
\label{W3A}
\\
T_1&= (x^2+x_1^2+1)(1-x_1)(x_1-x)+{\cal O}(\epsilon),
\\
T_2&=  \bigl( 1+x_1^2 +\epsilon(1-x_1)^2\bigr)
      \bigl( x^2+x_1^2 +\epsilon(x_1-x)^2\bigr),
\label{T2A} 
\\
T_3&= x_1(x^2+x_1^2+1)+{\cal O}(\epsilon),
\\
q_1^2 &= -\frac{k_{1\perp}^2}{\alpha_1} = -\frac{q_{1\perp}^2}{\alpha_1}.
\end{align}
As before, we will calculate only the difference w.r.t.\ the result with
cut-off on the virtuality, $-q^2<Q^2$. Therefore, the pole coming from $1/q^2$
integrand is eliminated and we are forced to keep only terms that generate
the $\epsilon$ pole from the $dk_{1\perp}^2$ integral. This means that we keep only
$T_2$,  set to zero all other $\epsilon$-terms and
expand $dk_{1\perp}^2$-integral, i.e.\
\begin{align}
&T_1=T_3=0,
\notag\\
&\epsilon\to 0 \;\;\;\hbox{except } k_{1\perp}^{2\epsilon},
\label{condi}
\\
\notag
&\int dk_{1\perp}^2 k_{1\perp}^{-2+2\epsilon} \to
\frac{1}{\epsilon}\int\ \delta(k_{1\perp}^2) dk_{1\perp}^2.
\end{align}
This way we obtain
\begin{align}
\Gamma_{Br}^{(q)} =&\; C_F^2 \frac{g^4}{(2\pi)^6} 4\hbox{PP} \Biggl[
  ~\int\limits_{-q^2>Q^2}
        \frac{d\alpha_2}{2\alpha_2^3}
         d^{2}\vec k_{2\perp} 
  \int \frac{d\alpha_1}{2\alpha_1}  d^{2+2\epsilon}\vec k_{1\perp}
  \delta_{1-x-\alpha_1-\alpha_2}
  \frac{k_{2\perp}^2}{q^4}
      \frac{1}{k_{1\perp}^2}
   T_2(\epsilon=0)
                        \Biggr].
\label{ladderBB}
\end{align}

The matching counter term $\Gamma_{Br}^{Ct}$ differs only by the ``split"
of the trace $W_{Br}^{ct}$ and an additional projection operator. The projection
operator
performs two actions: picks the $\epsilon$-poles and sets on-shell the incoming
quark ($q_1$ in our case). These are minor modifications to (\ref{ladderA},
\ref{W3A}):
\begin{align}
\Gamma_{Br}^{Ct} =&\; C_F^2 \frac{g^4\;x}{(2\pi)^6} \hbox{PP} \Biggl[
  \frac{(2\pi)^{-2\epsilon}}{\mu^{2\epsilon}}
  \int\limits_{-q^2>Q^2}
        \frac{d\alpha_2}{2\alpha_2}
         d^{2+2\epsilon}\vec k_{2\perp} 
  \frac{1}{q^4} W_{Br2}
             \bigg|_{q_1^2=0}
\notag\\ &\times
\hbox{PP}\Biggl(
\frac{(2\pi)^{-2\epsilon}}{\mu^{2\epsilon}}   
  \int \frac{d\alpha_1}{2\alpha_1}  d^{2+2\epsilon}\vec k_{1\perp}
      \frac{\alpha_1^2}{k_{1\perp}^4}  W_{Br1}
  \delta_{1-x-\alpha_1-\alpha_2}
                     \Biggr) \Biggr],
\end{align}
where
\begin{align}
W_{Br2}=&
\frac{1}{4qn} \hbox{Tr} \Bigl(
  \hat n \hat q \hat \gamma^\mu \hat q_1 \hat \gamma^\nu \hat q 
           \Bigr)  {d_{\mu\nu}(k_2)_{\Bigl|}}_{q_1^2=0}
=
-2q^2\frac{1}{x\alpha_2}
      (x_1^2+x^2 +\epsilon(x_1-x)^2),
\\
W_{Br1} =&
\frac{1}{4q_1n} \hbox{Tr} \Bigl(
  \hat n \hat q_1 \hat \gamma^\alpha \hat p \hat \gamma^\beta \hat q_1
           \Bigr) d_{\alpha\beta}(k_1)
=
-2q_1^2\frac{1}{x_1\alpha_1}
      (1+x_1^2+\epsilon(1-x_1)^2),
\end{align}
and thanks to the condition $q_1^2=0$:
\begin{align}
q_1^2 = -\frac{k_{1\perp}^2}{\alpha_1},
\;\;\;\;
q^2 {{\Bigl|}_{q_1^2=0}}=
-x\biggl(\frac{k_{1\perp}^2}{\alpha_1}+\frac{k_{2\perp}^2}{\alpha_2}
        \biggr) -k_\perp^2{{\Bigl|}_{k_{1\perp}^2=0}}
  = - \frac{x_1 k_{2\perp}^2}{\alpha_2}.
\label{lotraceA}
\end{align}
We obtain
\begin{align}
\Gamma_{Br}^{Ct} 
=&\; C_F^2 \frac{g^4}{(2\pi)^6} 4 \hbox{PP} \Biggl[
  ~\int\limits_{-q^2>Q^2}
        \frac{d\alpha_2}{2\alpha_2^3}
         d^{2}\vec k_{2\perp} 
  \frac{k_{2\perp}^2}{q^4}
             \bigg|_{q_1^2=0}
  \int \frac{d\alpha_1}{2\alpha_1}  d^{2+2\epsilon}\vec k_{1\perp}
      \frac{1}{k_{1\perp}^2}
  \delta_{1-x-\alpha_1-\alpha_2}
                        T_2(\epsilon=0)
\Biggr].
\label{ladderCT}
\end{align}
It is easy to verify now that these two quantities,
$\Gamma_{Br}$ and $\Gamma_{Br}^{Ct}$, are identical under the conditions
(\ref{condi}) and the net change of the kernel is zero.

In  Appendix~A we evaluate the change of the ladder graph alone caused by the
change of cut-off. This quantity is of interest for example in the construction
of Monte Carlo algorithms.

\section{Conclusions}
In this paper we have discussed the change of the DGLAP kernel $P_{qq}$
due to the change
of the evolution variable. We have demonstrated that at the NLO level majority
of the choices of the evolution variables lead to the same kernel, but there are
ones, like maximal transverse momentum, that correspond to modified kernel. We
have shown the mechanism responsible for the change and we have formulated a simple rule
to identify classes of variables that leave the kernel unchanged at the NLO
level.

There is an important open question related to our analysis: is
the kernel dependence specific to the CFP method and specifically to the
presence of the geometrical cut-off $\delta$? If all the singularities,
including the ``spurious" ones, were regulated by the dimensional
regularization, the structure of the $\epsilon$ poles would be reacher, more 
graphs would have higher-order poles in $\epsilon$ and would contribute to
the modification of the kernel. This would, however, be a surprising result
showing that the choice of the seemingly dummy technical regulator has a
physical consequences. The same question holds for the modification of the
original PV prescription of \cite{Curci:1980uw} to the NPV one used in this
note.

\section*{Acknowledgments}
This work is partly supported by 
 the Polish National Science Center grant DEC-2011/03/B/ST2/02632
 and
 the Polish National Science Centre grant UMO-2012/04/M/ST2/00240.

\appendix
\section{Change of ladder graph with cut-off}
In the appendix we calculate change
of the $\Gamma_{Br}$ for various cut-offs as it can be useful in constructing
MC algorithms. Let us continue with eq.\ (\ref{ladderA}) and let us implement
the conditions (\ref{condi}):
\begin{align}
& \int d^{2+2\epsilon}\vec k_{1\perp} \frac{1}{k_{1\perp}^{2}}=
\int \frac{1}{2} \frac{dk_{1\perp}^2}{k_{1\perp}^2} k_{1\perp}^{2\epsilon}
d\Omega_{1+\epsilon}^{(k_{1\perp})}
\to
\int \frac{1}{2} dk_{1\perp}^2
\frac{1}{\epsilon}\delta(k_{1\perp}^2)
d\Omega_{1}^{(k_{1\perp})}
=
2\pi \frac{1}{2\epsilon}
\label{condi:del}
\\
& \int_L^U d^{2+2\epsilon}\vec k_{2\perp} \frac{1}{k_{2\perp}^{2}}
\to
\int_L^U \frac{1}{2} dk_{2\perp}^2 k_{2\perp}^{-2}
d\Omega_{1}^{(k_{2\perp})}
=
\pi \ln\frac{U}{L}.
\end{align}
The lower limit on integral $d^{2+2\epsilon}\vec k_{2\perp}$ follows from the
fact that we compute the difference w.r.t.\ the virtuality-based formula. This
leads to the condition 
\begin{align}
Q^2 < -q^2 = \frac{x_1 k_{2\perp}^2}{\alpha_2}
\to
k_{2\perp}^2> Q^2 \frac{\alpha_2}{x_1}.
\end{align}
The upper limit depends on the chosen evolution variable. We will examine a few
cases. The cut-offs and their simplified version once the condition
(\ref{condi:del}), i.e.\ $k_{1\perp}=0$, is applied are as follows:
\begin{align}
 \left.
\begin{array}{l}
(A):~~\max\{k_{1\perp},k_{2\perp}\} 
\\
(B):~~k_{1\perp}+k_{2\perp}
\\
(C):~~\max\bigl\{\frac{k_{1\perp}}{\alpha_1},
                 \frac{k_{2\perp}}{\alpha_2}\bigr\} 
\\
(D):~~ \frac{|\vec k_{1\perp}+\vec k_{2\perp}|}{\alpha_1+\alpha_2} 
\end{array} 
\right\}
\overset{k_{1\perp}=0}{\Longrightarrow}
\left\{
\begin{array}{l}
(A):~~k_{2\perp}<Q
\\
(B):~~k_{2\perp}<Q
\\
(C):~~{k_{2\perp}}<{\alpha_2}Q
\\
(D):~~k_{2\perp}<{(1-x)}Q
\end{array} \right.
\end{align}
Eq.\ (\ref{ladderA}) transforms now into
\begin{align}
\Delta\Gamma_{Br}^{U-q^2} =&\; C_F^2 
          \Bigl(\frac{\alpha}{2\pi}\Bigr)^2
                     \Biggl[
  \int
        \frac{d\alpha_2}{\alpha_2}
         \ln\frac{U}{L}
  \int \frac{d\alpha_1}{\alpha_1}  \frac{1}{2\epsilon}
  \delta_{1-x-\alpha_1-\alpha_2}
\frac{1}{x_1^2}
     \bigl( 1+x_1^2 \bigr) \bigl( x^2+x_1^2 \bigr)
                     \Biggr].
\label{ladderB}
\end{align}
Let us continue with each case separately.
\\ \\
{\bf Cases (A) and (B):  $\max\{k_{1\perp},k_{2\perp}\}$ and $k_{1\perp}+k_{2\perp}$}
\begin{align}
\Gamma_{Br}^{k_\perp-q^2} =&\; C_F^2 
          \Bigl(\frac{\alpha}{2\pi}\Bigr)^2
                 \frac{1}{2\epsilon}   
  \int\limits_0^{1-x} \frac{d\alpha_1}{\alpha_1\alpha_2}  
\frac{1}{x_1^2}
     \bigl( 1+x_1^2 \bigr) \bigl( x^2+x_1^2 \bigr)
         \ln\frac{x_1}{\alpha_2}
\\=&\;\notag
 C_F^2 \Bigl(\frac{\alpha}{2\pi}\Bigr)^2 \frac{1}{2\epsilon}\frac{1}{1-x} 
  \int\limits_0^{1-x} {d\alpha_1}
(U_0 +U_l +U_u),
\\ 
U_0 =&\;
  \Bigl(\frac{1}{1-x_1}+\frac{1}{x_1-x}\Bigr)
     \frac{1}{x_1^2} \bigl( 1+x_1^2 \bigr) \bigl( x^2+x_1^2 \bigr)
         \ln\frac{x_1}{x_1-x}
- U_l -U_u ,
\\ 
U_l =&\;
\frac{1}{1-x_1}2(1+x^2) \ln\frac{1}{1-x},
\\ 
U_u =&\;
\frac{1}{x_1-x}2(1+x^2) \ln\frac{x}{x_1-x},
\label{ladderC}
\end{align}
where we have subtracted and added the singular integrals of the  $I_{0,1}$ type. Direct
integration gives%
\begin{align}
\int\limits_0^{1-x} {d\alpha_1} U_0 =&\;
-(1-x)^2 +(1+x^2)\ln^2x +(1+3x^2)\frac{\pi^2}{6} +2(1-x)^2\ln(1-x) 
\notag \\ &
-(x^2-1) \Li_2(x) +x(1-x)\ln x
\\
\int\limits_0^{1-x} {d\alpha_1} U_l =&\;
2(1+x^2) (I_0+\ln(1-x))\ln\frac{1}{1-x} 
\\
\int\limits_0^{1-x} {d\alpha_1} U_u =&\;
2(1+x^2) (I_0+\ln(1-x))\ln{x}
-
2(1+x^2) \Bigl(I_1+\frac{1}{2}\ln^2(1-x)\Bigr)
\notag
\\ =& \;
2(1+x^2) \Bigl(-I_1^{(1-x)}+I_0^{(1-x)}\ln\frac{x}{1-x}\Bigr),
\end{align}
where
\begin{equation}
\begin{split}
I_0^{(1-x)} &= I_0 + \ln(1-x),\\
I_1^{(1-x)} &= I_1 - I_0\ln(1-x) + \frac{1}{2}\ln^2(1-x).
\end{split}
\end{equation}
Hence
\begin{align}
\Gamma_{Br}^{k_\perp-q^2} =&\; 
C_F^2 \Bigl(\frac{\alpha}{2\pi}\Bigr)^2 \frac{1}{2\epsilon}
\biggl[
-(1-x) -(1+x) \frac{\pi^2}{6}
       +2(1-x)\ln(1-x) +(1+x) \Li_2(x) 
\notag \\ & \;
+x\ln x
+2\frac{1+x^2}{1-x} \Bigl(-I_1^{(1-x)}+I_0^{(1-x)}\ln\frac{x}{(1-x)^2}
                +\frac{1}{2}\ln^2x +\frac{\pi^2}{6}\Bigr)
\biggr].
\end{align}
\\ 
{\bf Case (C):  $\max\bigl\{\frac{k_{1\perp}}{\alpha_1},
                 \frac{k_{2\perp}}{\alpha_2}\bigr\}$}
\begin{align}
\Gamma_{Br}^{k_\perp/a-q^2} =&\; C_F^2 
          \Bigl(\frac{\alpha}{2\pi}\Bigr)^2
                 \frac{1}{2\epsilon}   
  \int\limits_0^{1-x} \frac{d\alpha_1}{\alpha_1\alpha_2}  
\frac{1}{x_1^2}
     \bigl( 1+x_1^2 \bigr) \bigl( x^2+x_1^2 \bigr)
         \ln({x_1}{\alpha_2})
\\=&\;\notag
C_F^2 \Bigl(\frac{\alpha}{2\pi}\Bigr)^2 \frac{1}{2\epsilon}
\biggl[
1-x +(1+x)\ln^2x +(1+x)\frac{\pi^2}{6} 
-2(1-x)\ln(1-x) 
-(2-x)\ln x
\notag \\ &
 -(1+x) \Li_2(x)
+2\frac{1+x^2}{1-x} \Bigl(I_1^{(1-x)}+I_0^{(1-x)}\ln\bigl(x(1-x)^2\bigr)
               -\frac{\pi^2}{6} +\frac{1}{2}\ln^2x
          \Bigr)
\biggr].
\end{align}
\\ 
{\bf Case (D):  ${|\vec k_{1\perp}+\vec
k_{2\perp}|}/(\alpha_1+\alpha_2) $}
\begin{align}
\Gamma_{Br}^{k_\perp/(1-x)-q^2} =&\; C_F^2 
          \Bigl(\frac{\alpha}{2\pi}\Bigr)^2
                 \frac{1}{2\epsilon}   
  \int\limits_0^{1-x} \frac{d\alpha_1}{\alpha_1\alpha_2}  
\frac{1}{x_1^2}
     \bigl( 1+x_1^2 \bigr) \bigl( x^2+x_1^2 \bigr)
         \ln\frac{x_1(1-x)}{\alpha_2}
\notag
\\=&\;\notag
C_F^2 \Bigl(\frac{\alpha}{2\pi}\Bigr)^2 \frac{1}{2\epsilon}
\biggl[
 -(1-x) -(1+x)\Li_2(1-x)+x\ln x
\notag \\ & \;
2\frac{1+x^2}{1-x} \Bigl(-I_1^{(1-x)}+I_0^{(1-x)}\ln{x}
               +\frac{\pi^2}{6} +\frac{1}{2}\ln^2 x\Bigr)
\biggr].
\end{align}

\bibliographystyle{elsarticle-num}
\bibliography{radcor}

\end{document}